\documentclass[ twocolumn,preprintnumbers,amsmath]{revtex4}
\usepackage{epsfig}
\usepackage{bm}
\usepackage{amsmath}

\usepackage{amssymb}

\begin{document}
\title{Comment on "Negative heat capacities 
and first order phase transitions in
nuclei" by L.G. Moretto et al  }
\author{ F. Gulminelli} 
\affiliation{LPC Caen (IN2P3 - CNRS / EnsiCaen et Universit\'{e}), F-14050
Caen C\'{e}dex, France }

\author{Ph.Chomaz}
\affiliation{GANIL ( DSM - CEA / IN2P3 - CNRS),
B.P.5027, F-14021 Caen C\'edex, France}

\begin{abstract}
In a recent paper L.G. Moretto et al \cite{Moretto}
claim that the negative heat capacities presented in our 
previously published paper \cite{PRL2000} are
"artifacts" coming from the use of periodic boundary 
conditions in the Lattice-Gas calculations. We stress 
in this comment that this claim is wrong: in ref. \cite{PRL2000}
we did {\it not} use periodic boundary conditions and anyhow the boundary 
conditions are irrelevant for the statistical ensemble used in 
\cite{PRL2000}.
The second claim of \cite{Moretto} is 
that, because of the Coulomb repulsion, systems 
"with $A>60$ should present no anomalous
negative heat capacities". We show that this conclusion  
is contradicted by exact Lattice-gas simulations including
Coulomb forces which present negative heat capacities
even for $A>200$. 
\end{abstract}
\maketitle   
 
Let us start with the discussion about the boundary conditions. 
It is clearly stated in  ref. \cite{PRL2000} 
that"$L^3$ [the lattice size] is 
large enough (typically greater than $20^3$ lattice 
sites) so that the boundary conditions do not affect 
the calculations with a constraining $\lambda$." This
means that the results of ref.  
\cite{PRL2000} do not depend upon the conditions used
at the boundary, and in fact the calculations were made
{\it without} periodic boundary conditions. At that time
we checked the independence of the boundaries 
comparing $N=8000$ and $N=27000$ lattices.
To see a sizeable effect of the boundary, we have
decreased the size of the lattice.  
A fast calculation of $50000$ events  
for $216$ particles in a $N=5832$ lattice
at an energy $E=0.4\epsilon $ 
confined by a Lagrange multiplier $\log \lambda=-8 $ , 
gives a temperature $T=(0.687 \pm 0.004)\epsilon$, 
a heat capacity $C = -16.3 \pm 0.2$, 
and a kinetic energy fluctuation $\sigma_k^2/T^2 =1.64 \pm 0.03$.
If periodic boundary conditions are imposed, the temperature
becomes $T=(0.682 \pm 0.004)\epsilon$, 
the heat capacity $C = -16.9 \pm 0.4 $ 
and the fluctuation $\sigma_k^2/T^2 = 1.63 \pm 0.07$. 
The temperature decrease in the phase transition region is $\Delta T=(7.6 \pm
0.4)10^{-3} \epsilon$ ($\Delta T=(6.7 \pm 0.6)10^{-3} \epsilon$) with
(without) periodic boundary conditions.
This  shows that
the boundary conditions do not affect the thermodynamics.
Indeed, in ref. \cite{PRL2000} we have analyzed 
an ensemble of particles 
in the vacuum for 
which only the average spatial extension of the system
is defined (isobar ensemble). Then, provided that the lattice in which the system 
has still to be discretized
for technical reasons is big enough, 
boundary conditions are irrelevant since the particles never explore 
the outer region. 

In a previous paper \cite{PRL1999} 
we have considered a canonical 
Lattice Gas model in a box of constant volume (isochore 
ensemble), and there we have used periodic boundary conditions.  
In these calculations negative compressibilities 
are reported but the heat capacity is always positive.
In fact, 
in the canonical ensemble the heat 
capacity is proportional to the energy variance 
and can never be negative \cite{Shro}. 

Concerning the isochore microcanonical ensemble,
the published results \cite{Richert} show that
the heat capacity 
at constant volume $C_V$ is always positive.
Therefore, contrary to what is claimed in ref. \cite{Moretto}, 
published results dealing with Lattice Gas calculations 
with typical nuclear size systems and periodic boundary conditions 
\cite{Richert,IMFM} do not show any negative heat capacity.  

We may incidentally note that recently 
a debate has started 
in the statistical physics community after the observation of a negative 
heat capacity branch for very large systems in 
the the constant-magnetization microcanonical  
Ising model \cite{Pliemling} (equivalent to the 
microcanonical isochore Lattice Gas model). 
In these cases however, the densities involved are 
so low 
that once again boundary conditions are irrelevant. 
Moreover, the sizes
discussed are orders of magnitude larger than the nuclear ones.

Finally we would like to stress that we have recently demonstrated
that negative heat capacities in finite systems 
are the origin of first order phase 
transitions with a finite latent heat: 
to present an energy discontinuity 
at the thermodynamic limit, 
finite systems {\it must} present a negative microcanonical 
heat capacity if the number 
of particles is large enough \cite{zeroes}. 

\begin{figure}[tbp]
\includegraphics[width=.9\linewidth]{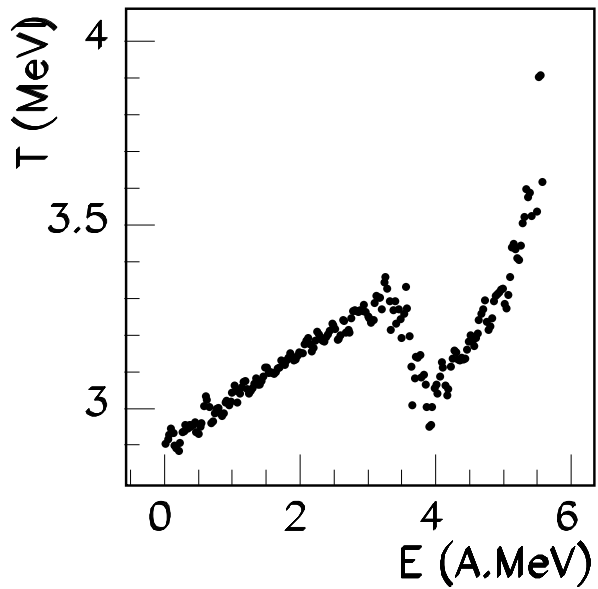}
\caption{ Caloric curve for the Lattice Gas hamiltonian
augmented with the Coulomb interaction in the microcanonical isobar ensemble
($\log \lambda = -7, A=207, Z=82$) from ref.\protect\cite{vero}. The coupling
between neighboring sites is fixed to $\epsilon = 5.5 MeV$.   }
\end{figure}

Let us now comment about the effect of Coulomb.  
The conclusion of ref.\cite{Moretto} is 
based on a simple model of a unique cluster in 
equilibrium with a gas. However, 
this configuration is 
not  the most probable 
one both in statistical models \cite{smm,gross} and in nuclear physics 
experiments at high excitation energy  \cite{data}. 
In particular, the negative heat capacity 
region is characterized by a large ($m \geq 3$) 
multiplicity of fragments \cite{palluto}. 
Therefore any conclusion drawn considering only the specific channel of
multiplicity 1 is not valid. 
If we would follow the simple model of ref. \cite{Moretto} 
and consider partitions with $m=3-5$ fragments of similar mass
\cite{data}, 
this would lead to a figure analogous to 
figure 5 of ref. \cite{Moretto} but with an abscissa approximately 
multiplied by  $m$ (or more, since all the light particles have to be included),
because the inter-fragment coulomb interaction  
can be neglected if the volume is large. 
One would then find $A \approx 200$ as 
the limit of negative heat capacities. 

Of course fragments do not have the same size and 
to get a quantitative result 
one needs to compute the relative weight of all the possible 
channels; this is indeed what is done in nuclear statistical models
\cite{gross,smm,raduta} which show that the intermediate state between 
the compound nucleus and the vaporized system is 
multifragmentation: the presence of many drops in equilibrium.
 
In the case of the Lattice Gas model the different partitions 
and their relative
weight can be calculated without any approximation.  
In the negative heat capacity energy regime there are in average 
three fragments of size greater than 4 \cite{PRL2000}.
The sudden opening of this multifragment channel\cite{gross} 
causes a convex intruder in the entropy that is responsible 
of the negative heat capacity. 
The effect of the Coulomb interaction in the Lattice Gas model 
can be appreciated from  
oFigure 1 which shows a caloric curve at constant $\lambda=\beta p$
for a system of $A=207$ particles and charge $Z=82$ \cite{vero}.
A clear backbending is visible. 
In the charged Lattice-gas model, 
the energy interval corresponding to the backbending
decreases with increasing charge \cite{vero}, 
showing that the idea  that Coulomb
tends to suppress the negative heat capacity 
is qualitatively correct.
However, when all the available channels are 
correctly weighted, 
the quantitative effect is very different
then the one reported in ref. \cite{Moretto}.
Only a slight effect of the Coulomb interaction has
also been reported in the isochore framework \cite{carmona}.

Obviously the result of Figure 1 is model dependent.
Different macroscopic models show a higher \cite{raduta} or lower \cite{smm}
sensitivity to the Coulomb depending on the detailed implementation 
of the Coulomb and nuclear interactions.
On the other hand molecular dynamics (which can be solved exactly as the Lattice
Gas model) leads to caloric curves which are almost independent of Coulomb
\cite{claudio}. Whether this can be interpreted as an evidence of metastable
long time tails, the relaxation time of long range interactions being
excessively long, is a subject of debate \cite{tsallis}.   
In any case these examples and the results of Figure 1 show that 
a definitive understanding of the effect of a non saturating long range
interaction on a first order phase transition cannot certainly be achieved
through the oversimplified model of ref. \cite{Moretto}.

\end{document}